\documentclass[twocolumn,showpacs,prb]{revtex4}
\usepackage{epsfig}
\newcommand{\vk}{{\mbox{\boldmath$k$}}}

\newcommand{\vsk}{{\mbox{\boldmath$k$}}}
\newcommand{\vg}{\mbox{\boldmath$g$}}
\newcommand{\hvg}{\hat{\mbox{\boldmath$g$}}}
\newcommand{\vsig}{\mbox{\boldmath$\sigma$}}
\newcommand{\vd}{\mbox{\boldmath$d$}}
\newcommand{\mhx}{\hat{\mbox{\boldmath$x$}}}
\newcommand{\mhy}{\hat{\mbox{\boldmath$y$}}}
\newcommand{\mhz}{\hat{\mbox{\boldmath$z$}}}
\newcommand{\vh}{\mbox{\boldmath$h$}}
\newcommand{\vH}{\mbox{\boldmath$H$}}

\begin{document}

\title{Superconductivty without inversion symmetry: MnSi versus CePt$_3$Si}
\author{P. Frigeri$^1$, D.F. Agterberg$^2$, A. Koga$^{1,3}$, and
  M. Sigrist$^1$}
\address{$^1$Theoretische Physik ETH-H\"onggerberg CH-8093 Z\"urich,
  Switzerland}
\address{$^2$Department of Physics, University of Wisconsin-Milwaukee,
  Milwaukee, WI 53201}
\address{$^3$Department of Applied Physics, Osaka University, Suita,
  Osaka 565-0871,Japan}

\begin{abstract}
Superconductivity in materials without spatial inversion symmetry
is studied. We show that in contrast to common believe,
spin-triplet pairing is not entirely excluded in such systems.
Moreover, paramagnetic limiting is analyzed for both spin-singlet
and triplet pairing. The lack of inversion symmetry reduces the
effect of the paramagnetic limiting for spin-singlet pairing.
These results are applied to MnSi and CePt$_3$Si.
\end{abstract}


\maketitle

Cooper pairing in the spin-singlet channel relies on the presence
of time reversal symmetry (Anderson's theorem); the paired
electron states are related by time reversal and are consequently
degenerate \cite{and59}. If this degeneracy is lifted, for
example, by a magnetic field or magnetic impurities coupling to
the electron spins, then superconductivity is weakened or even
suppressed. For spin-triplet pairing, Anderson noticed that
additionally inversion symmetry is required to obtain the
necessary degenerate electron states \cite{and84}. Consequently,
it became a widespread view that a material lacking an inversion
center would be an unlikely candidate for spin-triplet pairing.
For example, the absence of superconductivity in the paramagnetic
phase of MnSi close to the quantum critical point to itinerant
ferromagnetism was interpreted from this point of view
\cite{mat98,sax00}. Near this quantum critical point the most
natural spin fluctuation mediated Cooper pairing would occur in
the spin-triplet channel. However, MnSi has the so-called B20
structure (P2$_1$),
without inversion center, inhibiting spin-triplet pairing.

Recently, superconductivity has been discovered
in the heavy fermion compound CePt$_3$Si, another
system without inversion symmetry (P4mm)\cite{bau03} . The upper
critical field $ H_{c2} $ exceeds the usual
paramagnetic limiting field, which might indicate that here
nevertheless spin-triplet pairing is realized.
Since there is no experimental
information on the pairing symmetry in this material so far, it is
worth examining the options for Cooper pairing in this case.

The aim of this letter is to discuss two points for time-reversal
invariant materials without inversion centers. The first is
concerned with the possible existence of spin-triplet pairing. The
second addresses the problem of paramagnetic limiting
(Clogston-Chandrasekar-Pauli limiting). The result of this
discussion will  be applied to the two materials mentioned above:
MnSi and CePt$_3$Si.

{\it Model:} We use a single-band model with electron band energy
$ \xi_{\vsk} $ measured from the Fermi energy where electrons with
momentum $ \vk $ and spin  $ s$ are created (annihilated) by the
operators $ c^{\dag}_{\vsk s} $ ($ c_{\vsk s} $). The Hamiltonian
including the pairing interaction is
\begin{equation}
H = \sum_{\vsk,s} \xi_{\vsk} c^{\dag}_{\vsk s} c_{\vsk s} +
\frac{1}{2} \sum_{\vsk, \vsk'}\sum_{s,s'}V_{\vsk, \vsk'}
c^{\dag}_{\vsk s} c^{\dag}_{-\vsk s'} c_{- \vsk' s'} c_{\vsk' s}.
\end{equation}
This system possesses time reversal and inversion symmetry ($
\xi_{\vsk} = \xi_{- \vsk} $) and the pairing interaction does not
depend on the spin and favors either even-parity (spin-singlet) or
odd-parity (spin-triplet) pairing as required. Following the
standard weak-coupling approach, we define the interaction to be
finite and attractive close to the Fermi energy with the cutoff
energy $ \epsilon_c$, and to depend on the momenta only through
the angular dependence. The absence of inversion symmetry is
introduced by an additional term, $ H_p$, to the Hamiltonian which
removes parity but conserves time reversal symmetry, {\it i.e.} $
I H_p I^{-1} = - H_p $ and $ T H_p T^{-1} = H_p $. We can write
such a single-particle term as
\begin{equation}
H_p = \alpha \sum_{\vsk, s, s'} \vg_{\vsk} \cdot \vsig_{ss'}
c^{\dag}_{\vsk s} c_{\vsk s'}
\label{hcorr}
\end{equation}
where $ \vsig $ denotes the Pauli matrices and $ \vg_{-\vsk} = -
\vg_{\vsk} $ (this satisfies the above condition since $ I \vsig
I^{-1} = \vsig $ and $ T\vsig T^{-1} = - \vsig $). It is
convenient to normalize $ \vg_{\vsk} $ so that the average over
the Fermi surface $ \langle \vg_{\vsk}^2 \rangle_{\vsk} =1 $, in
the numerical calculations we will impose this constraint.
We will keep $\vg_{\vsk}$ arbitrary
and later provide a specific form of $ \vg_{\vsk} $ for MnSi and
CePt$_3$Si. The normal state Green's function becomes,
\begin{equation}
G^0(\vk, i \omega_n) = G_{+} (\vk, i \omega_n) \sigma_{0} +
  \hvg_{\vsk} \cdot   \vsig G_-(\vk, i \omega_n)
\end{equation}
where $ \sigma_0 $ is the unit matrix and
\begin{equation}
G_{\pm} (\vk, i \omega_n) = \frac{1}{2} \left[ (i \omega_n -
    \epsilon_{\vsk,+})^{-1}  \pm  (i \omega_n - \epsilon_{\vsk,-} )^{-1}
    \right] \; ,
\end{equation}
$ \epsilon_{\vsk,\pm} = \xi_{\vsk} \pm \alpha | \vg_{\vsk}| $ and $
\hvg_{\vsk} = \vg_{\vsk} / | \vg_{\vsk} | $ ($ | \vg | =
\sqrt{\vg^2} $) \cite{edelstein95}. The Fermi surface splits into two
sheets with different spin structure. These two sheets touch whenever ${\bf
g}({\bf k})=0$.

{\it Superconducting instability:} We now use the BCS decoupling
scheme and determine the linearized gap equation in order to
calculate the transition temperature $ T_c $:
\begin{widetext}
\begin{equation}
\Delta_{ss'} (\vk) = - k_B T \sum_{n, \vsk'} \sum_{s_1,s_2}
V_{\vsk,\vsk'} G^0_{ss_1}(\vk', i\omega_n)  \Delta_{s_1,s_2}
(\vk') G^0_{s's_2} ( - \vk', - i\omega_n)
\end{equation}
The gap function is decomposed into a spin-singlet [$ \psi(\vk) $]
and a triplet [$ \vd(\vk) $] part, $\Delta(\vk) = \{ \psi(\vk)
\sigma_0 + \vd (\vk) \cdot \vsig \} i \sigma_y $.  For simplicity
we assume that the gap functions have the same magnitude on both
Fermi surface sheets. This allows us to write the linearized gap
equations as
\begin{equation}
\psi (\vk) = - k_B T \sum_{n, \vsk'} V_{\vsk, \vsk'} \Big\{
[G_+ G_+ + G_- G_- ]\psi (\vk')  + [G_+ G_- + G_- G_+] \hvg_{\vsk'} \cdot \vd ( \vk') \Big \}
\end{equation}
and
\begin{equation}
\vd (\vk) = - k_B T \sum_{n, \vsk'} V_{\vsk, \vsk'} \Big\{
[G_+ G_+ +
 G_- G_- ] \vd (\vk')  + 2 G_- G_- [\hvg_{\vsk'} (\hvg_{\vsk'}
\cdot \vd ( \vk') ) - \vd(\vk') ]
+ [G_+ G_- + G_- G_+] \hvg_{\vsk'} \psi(\vk') \; ,
\Big\}
\end{equation}
\end{widetext}
where we have used the short notation for the products: $ G_a G_b =
G_a (\vk, i \omega_n) G_b (-\vk, - i \omega_n ) $ with $ a,b=\pm $.
For finite $ \alpha $, the spin-singlet and triplet channel are
coupled, an effect of the missing parity \cite{gor01}. However, this
coupling depends on the degree of particle-hole asymmetry or the
difference of the density of states on the two Fermi surface sheets,
which yields a coupling of the order $ \alpha / \epsilon_F  \ll
1 $. Thus, we ignore these coupling terms here and consider
the ``singlet'' and ``triplet'' channel of pairing separately.

For spin-singlet pairing we find that the transition temperature
($T_c$) is given by
\begin{equation}
\ln  \left( \frac{T_c}{T_{cs}}\right) = O \left(
\frac{\alpha^2}{\epsilon_F^2}  \right) \;.
\end{equation}
The transition temperature remains essentially unchanged from $ k_B T_{cs} =
\epsilon_c \exp(-1/\lambda_s) $ (here $T_{cs}$ is $T_c$ for
$\alpha=0$) with $ \lambda_s \psi(\vk) = N(0) \langle V_{\vsk,
\vsk'} \psi (\vk') \rangle_{\vsk'} $. For triplet pairing the
equation for $T_c$ reads
\begin{equation}
\ln \left( \frac{T_c}{T_{ct}}\right) = 2 \langle \{ |\vd (\vk)|^2
- | \hvg_{\vsk}  \cdot \vd (\vk) |^2  \} f( \rho_{\vk})
  \rangle_{\vsk} + O \left(
  \frac{\alpha^2}{\epsilon_F^2} \right)
\label{tc-triplet}
\end{equation}
where $ k_B T_{ct} = \epsilon_c \exp(-1/ \lambda_t) $ with $
\lambda_t \vd(\vk) = N(0) \langle V_{\vsk, \vsk'} \vd (\vk')
\rangle_{\vsk'} $ and $ \rho_{\vsk} = \alpha | \vg_{\vsk} | / \pi
k_B T_c $. We use the normalized gap function with $ \langle | \vd
(\vk) |^2 \rangle_{\vsk} = 1 $ in all numerical calculations. The
function $ f (\rho) $ is defined as
\begin{equation}
f(\rho) = Re \; \sum_{n=1}^{\infty} \left( \frac{1}{2 n -1 + i \rho} -
\frac{1}{2n -1} \right) \; .
\label{tct}
\end{equation}
The correction term in Eq. (\ref{tc-triplet}) suppresses $ T_c $
in general. For a spherical Fermi surface and $ \alpha =0 $ all
gap functions with a given relative angular momentum $ \ell $ have
the same $ T_c $. Eq.(\ref{tct}) determines how this degeneracy is
lifted by the broken inversion symmetry. The highest $ T_c $ is
obtained for a state with $ \vd(\vk) \parallel \vg_{\vsk} $, for
which the right hand side of Eq.(\ref{tc-triplet}) vanishes and $
T_c = T_{ct} $. Hence we conclude that spin-triplet pairing is not
indiscriminately suppressed in the absence of an inversion center.
In principle, there may be spin-triplet pairing states which are
completely unaffected by the lack of inversion symmetry, taking
advantage of the spinor structure induced by $ \vg_{\vsk} $.

\begin{figure}
\epsfxsize=5 cm \center{\epsfbox{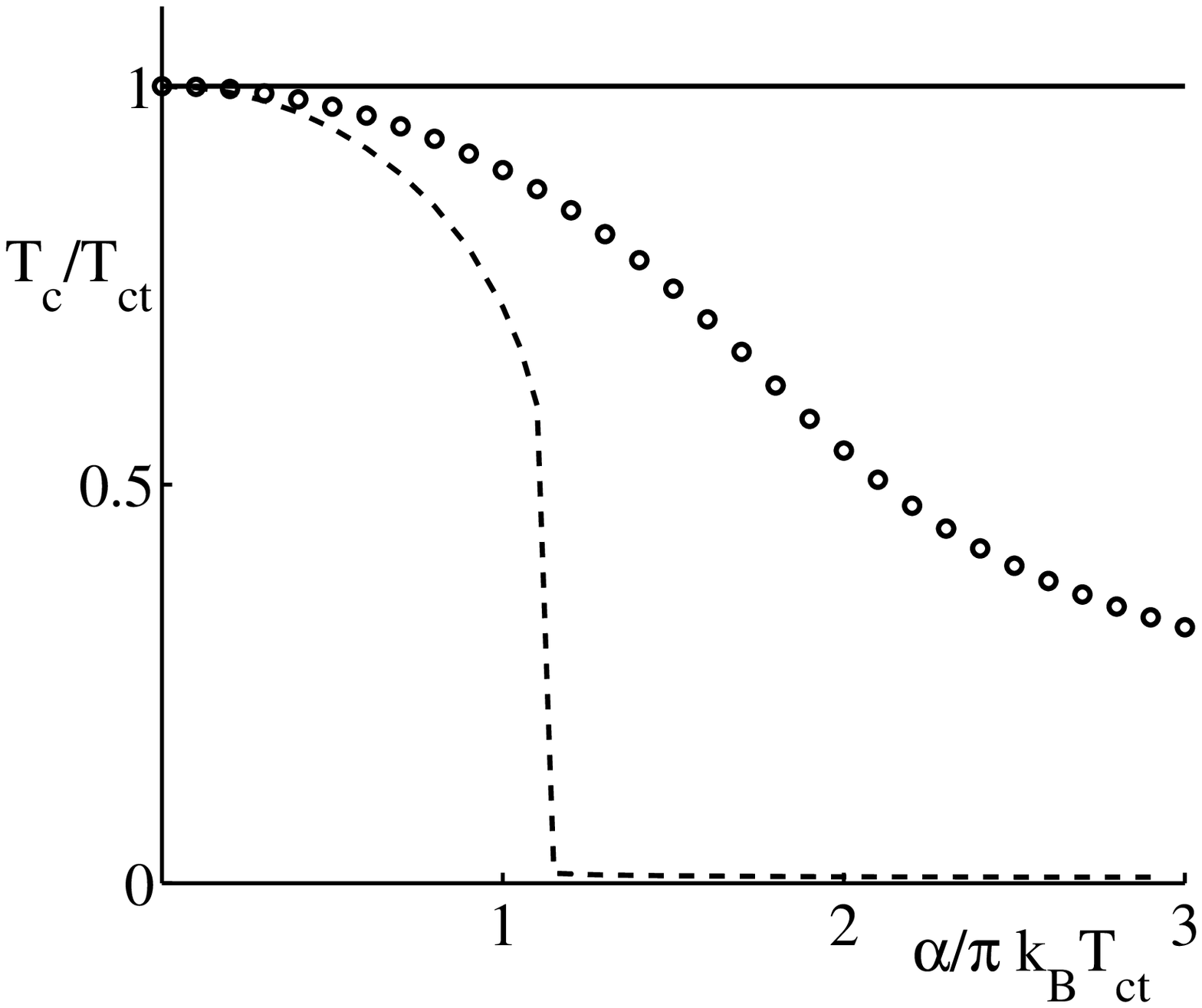}} \caption{Transition
  temperature as a function of $\alpha $ for $ \vg_{\vsk} = (-k_y, k_x
  ,0) $. The curves from top to bottom correspond to $\vd=\mhy k_x -\mhx k_y$,
  $\vd=\mhy k_x +\mhx k_y$, and $\vd=\mhx k_x +\mhy k_y+\mhz k_z$
  respectively.}
\label{fig1}
\end{figure}

{\it Structure of $\vg$:} The vector $ \alpha \vg_{\vsk} $ characterizes
and quantifies the absence of an inversion center in a crystal
lattice. In many cases the loss of an inversion center can be
viewed as  moving certain ions in the crystal lattice out of their
high-symmetry position. This gives rise to internal electric
fields that yield, through relativistic corrections, spin-orbit
coupling \cite{rashba60}. Furthermore, shifted ions can open new hopping
paths which involve atomic spin-orbit coupling on
intermediate ions.

We consider the form of $\vg_{\vsk}$ for our two examples: MnSi
and CePt$_3$Si. We start with the space group that corresponds to
the basic ``point group'' symmetry $ G $. Due to the lack of
inversion symmetry this group is reduced to a subgroup $ G' $. The
correction term $ \vg_{\vsk} \cdot \vsig $ is invariant under all
transformation of $ G' $, but not of $ G $. MnSi has the cubic
space group P2$_1$. The point group is only the tetrahedral group
$ T \in O_h $. The symmetry breaking term satisfying the above
conditions corresponds to the irreducible representation $ A_{2u}
$ of $ O_h $ which maps to $ A_1 $ of $ T_d $.
The expansion in $\vk$ leads to (we assume a spherical Fermi
surface for simplicity)
\begin{equation}
\vg_{\vsk} \cdot \vsig = k_y k_z (k_y \sigma_z - k_z \sigma_y) +
\mbox{cyclic perm. of} \; x,y,z
\end{equation}
which transforms like $ xyz$, a basis function of $ A_{2u} $ of $
O_h $.
The $ \vg $-vector
has 14 nodes on the Fermi surface, 6 along [100]- and 8
along [111]-directions. The $ \vd $-vector which remains unaffected by
the lack of inversion symmetry is parallel to $ \vg_{\vsk} $:
\begin{equation}
\vd (\vk) =
 \mhx k_x (k_y^2 - k_z^2) + \mhy k_y (k_z^2 -k_x^2) + \mhz k_z
( k_x^2 -k_y^2)
\end{equation}
which also belongs to $A_{2u} $, has
the same number of nodes as $ \vg $ and represents an $f$-wave
spin-triplet pairing state.

CePt$_3$Si is a tetragonal system with space group P4mm. Here the
removal of the inversion center leads to the point group  $
C_{4v} \in D_{4h} $ which corresponds to the loss of the basal
plane as a mirror plane ($ z \to - z $). One finds that $
\vg_{\vsk} \cdot \vsig = k_x \sigma_y - k_y \sigma_x  $,
which is a basis function of $
A_{2u} $ of $D_{4h} $. This term has the form of the well-known
Rashba spin-orbit coupling\cite{rashba60}. The $ \vg $-vector has only
two nodes
lying along the [001]-direction. In Fig.1 we show the reduction of
various spin-triplet pairing states with this $ \vg $-vector.
The favored pairing state is of $
p $-wave type: $ \vd (\vk) = \mhx k_y - \mhy k_x $ in $ A_{2u} $
of $ D_{4h} $, while other pairing states are severely
suppressed for $ \alpha > k_B T_c $.

With the reduced symmetry of a crystal, it is usually not possible
to find $ \vd(\vk) \parallel \vg_{\vsk} $ which satisfies the
linearized gap equation for a given pairing interaction $
V_{\vsk,\vsk'} $. Nevertheless, we could  determine a nearly
optimal spin-triplet state compromising between the pairing
interaction and the effect of $ H_p $. Very recently, Samokhin
{\it et al.} have carried out relativistic band structure
calculations which indicate that $ \alpha>>k_BT_c$ in CePt$_3$Si
\cite{samo03}. They also give a symmetry classification of the
possible pairing states \cite{samo03}.

The stability of the pairing state is not only decided by $ T_c $,
but also by the condensation energy; which is determined by the
shape of the quasi-particle gap in the weak coupling limit. The
Balian-Werthamer state (with a nodeless gap), $ \vd (\vk) = \mhx
k_x + \mhy k_y + \mhz k_z $,  is the most stable weak coupling
state in  a spherically symmetric (or cubic) system. However, in
the presence of broken inversion symmetry, Eq.(\ref{tc-triplet})
shows this state would generally have a lowered $ T_c $. Thus, for
small enough $\alpha$ ($\alpha<k_BT_{ct}$) there could be a second
superconducting phase transition below the onset of
superconductivity leading to a nodeless gap.

{\it Paramagnetic limiting:} Lifting the degeneracy of the spins
is detrimental to spin-singlet superconductivity, an effect known
as paramagnetic limiting. Spin-triplet pairing is less vulnerable
in this respect. In the absence of inversion symmetry, however,
this effect of pair breaking is modified. It is well-known that
impurity spin-orbit scattering reduces the effect of paramagnetic
limiting \cite{so-scat}. We show that an analogous effect occurs
in systems with broken inversion symmetry. For simplicity, we
ignore the effect of orbital pair breaking and include the
magnetic field only through its coupling to the spin. Our aim is
to demonstrate the effect of finite $ \alpha $ on the paramagnetic
limiting and an extended discussion for the upper critical field $
H_{c2} $ will be given elsewhere. We replace $ \alpha \vg_{\vk}
\to \alpha \tilde{\vg}_{\vk} = \alpha \vg_{\vk} - \vh $ with $ \vh
= \mu_B \vH $ (note: $ \tilde{\vg}_{-\vsk} \neq -
\tilde{\vg}_{\vsk} $). The linear gap equations yield the
transition temperatures for a continuous onset of
superconductivity \cite{comment1}. We first consider a dominant
spin-singlet pairing ($ \psi (\vk) $) and ignore the induced
spin-triplet pairing. Then the equation determining $T_c$ is
\begin{widetext}
\begin{equation}
{\rm ln} \left( \frac{T_c}{T_{cs}} \right) =\left\langle
|\psi(\vk)|^2 \left\{\left[f( \rho_{\vsk}^- )+f(
\rho_{\vsk}^+)\right]+ \frac{\alpha^2 \vg_{\vsk}^2 -
  \vh^2}{[(\alpha \vg_{\vsk} + \vh)^2 (\alpha \vg_{\vsk} - \vh)^2]^{1/2} }
  \left[f( \rho_{\vsk}^- )-f( \rho_{\vsk}^+)\right] \right\}\right\rangle_{\vsk}
  \label{para}
\end{equation}
\end{widetext}
with $ \rho_{\vk}^{\pm} = |\alpha \vg_{\vsk} + \vh |/ 2\pi k_B T_c
\pm|\alpha \vg_{\vsk} - \vh |/ 2\pi k_B T_c $. If it is possible
to choose $ \vh \perp \vg_{\vsk} $ for all $ \vk$ (as it is for
CePt$_3$Si), then in the small $T_c(h)/T_{cs}$ limit, the
paramagnetic limiting field obeys $h'^2\ln h'=-\alpha^2\ln
(T_c/T_{cs})$; with $ h'= |\vh|/\pi k_B T_{cs} $. In particular,
the paramagnetic limiting field {\it diverges} as $T\rightarrow 0$ (Fig.2).

For the spin-triplet channel we obtain analogously
\begin{widetext}
\begin{equation} \begin{array}{ll}
\displaystyle {\rm ln} \left( \frac{T_c}{T_{ct}} \right) =
&\displaystyle  \left\langle | \vd (\vk ) |^2 \left\{
\left[f(\rho_{\vk}^+ )+f(\rho_{\vk}^- )\right]+\frac{\alpha^2
\vg_{\vsk}^2 - \vh^2}{[ (
    \alpha \vg_{\vsk} + \vh)^2 ( \alpha \vg_{\vsk} - \vh)^2 ]^{1/2}}
 \left[
f(\rho_{\vk}^+ )-f(\rho_{\vk}^- )\right]\right\} \right\rangle_{\vsk} \\
& \displaystyle + 2 \left\langle \left[f(\rho_{\vsk}^+
)-f(\rho_{\vsk}^- ) \right]\frac{ | \vh \cdot
  \vd(\vk) |^2 -
  \alpha^2 | \vg_{\vsk} \cdot \vd(\vk) |^2}{[ (
    \alpha \vg_{\vsk} + \vh)^2 ( \alpha \vg_{\vsk} - \vh)^2 ]^{1/2}}
\right\rangle_{\vsk}
\end{array}
\label{d}
\end{equation}
\end{widetext}
For $ \alpha = 0 $ there is no paramagnetic limiting, provided $
\vd(\vk) \cdot \vh = 0 $ can be found for all $ \vk $. According
to Eq.(\ref{d}) paramagnetic limiting is absent, if for all $ \vk
$ $ \vh \perp \vd(\vk) $ and $ \vd(\vk)
\parallel \vg_{\vsk} $. For both the spin-singlet and spin-triplet
cases, finite ${\bf q}$
Fulde-Ferrell-Larkin-Ovchinnikov (FFLO) phases are often found
when $ \vh \cdot\vg_{\vsk}\ne 0 $ \cite{bar02,rpk03}. The role of
orbital effects on these phases is currently under investigation
\cite{rpk03}.

{\it Discussion of the two examples:} We start with MnSi, which
does not show superconductivity in the vicinity of the quantum
critical point of a ferromagnetic state. Given our result that
spin-triplet pairing is not suppressed completely by broken
inversion symmetry, it is useful to reexamine the reason for why
spin-triplet superconductivity is not observed. As one would
expect, the lack of inversion symmetry in this compound with
(cubic) B20-structure is crucial. According to our analysis the
pairing would have to occur in the $f$-wave channel in order to
survive the spin-orbit coupling effect. The fact that the strongly
anisotropic $f$-wave state is more difficult to stabilize by a
simple spin fluctuation mechanism than the $ p $-wave pairing
state might explains the absence of superconductivity in
MnSi.

\begin{figure}
\epsfxsize=7 cm \center{\epsfbox{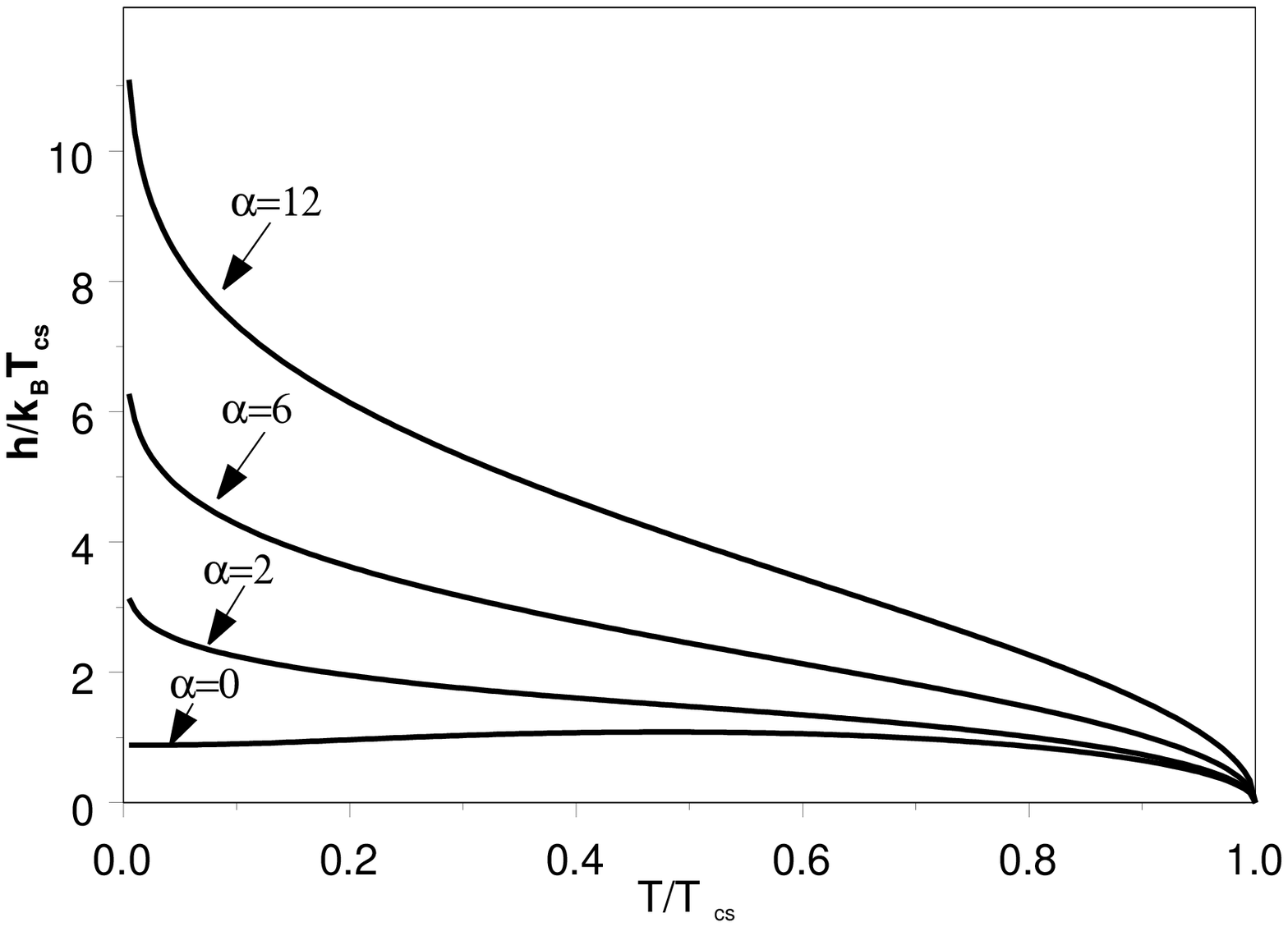}} \caption{Paramagnetic
limiting field for CePt$_3$Si for different $\alpha$ (in units
$k_BT_{cs}$). The field is applied along the four-fold symmetry
axis.} \label{fig2}
\end{figure}

Turning to CePt$_3$Si we may adopt two different points of view.
First, there is a protected $ p $-wave spin-triplet pairing state
($ \vd (\vk) = \mhx k_y - \mhy k_x $). This may indeed explain the
apparent absence of paramagnetic limiting observed in
polycrystaline samples\cite{bau03}. On the other hand, it is
important to notice that superconductivity appears here on the
background of antiferromagnetic order ($T_N \approx 2 K$), and it
seems more natural to assume a spin-singlet type of pairing. In this
case, we could argue that paramagnetic limiting for a singlet
state is rendered less effective by the presence of spin-orbit
coupling. To examine this possibility in more detail we have
determined the paramagnetic limiting field as a function of
$\alpha$ using Eq.~(\ref{para}) for the field along the four-fold
symmetry axis. This is shown in Fig.~\ref{fig2}, note that this
figure illustrates the divergent paramagnetic limiting field at
low temperatures described earlier. It would be very helpful to
study this system for single crystals, since for both the
spin-singlet and spin-triplet cases a large anisotropy in the
paramagnetic limiting field is predicted. The field along the
four-fold axis should give no paramagnetic limiting in both cases.
Moreover, the Knight shift should show related effects of the
spin-orbit coupling.

In conclusion, the analysis of the symmetry properties for the two
materials MnSi and CePt$_3$Si show that in the former system the
effect of the lack of inversion symmetry leads to more severe
restrictions for spin-triplet pairing than in the latter.
Furthermore, paramagnetic limiting for spin-singlet
superconductors is suppressed by broken inversion symmetry. In
many respects, CePt$_3$Si may become an ideal test system to study
the effect of missing inversion symmetry on the superconducting
phase \cite{edelstein95,gor01,yip02}.

{\it Acknowledgements:} We would like to thank E. Bauer, R.P.
Kaur, and T.M. Rice for stimulating discussions. This work was
supported by the Swiss National Science Foundation. DFA was also
supported by the National Science Foundation award DMR-0318665, an
award from Research Corporation, and the American Chemical Society
Petroleum Research Funds.

\eject




\noindent{\bf Errata on "Superconductivty without inversion
symmetry: MnSi versus CePt$_3$Si"} \vglue 0.5 cm

We have two corrections to make to our original paper. The first
involves a correction in the factor $\alpha_R$ in Fig.~2. The
$\alpha_R$ values in Fig.~2 should be divided by a factor
$\sqrt{3/2}$.

Also, we have omitted a contribution to the vector $\vg_{\vsk}$
for MnSi. Including this contribution gives
$\vg_{\vsk}=\alpha_1[k_x,k_y,k_z]+\alpha_2[k_x (k_y^2 - k_z^2),
k_y (k_z^2 -k_x^2),k_z ( k_x^2 -k_y^2)]$ (in the above paper we
have $\alpha_1=0$). The vector multiplied by $\alpha_1$ belongs to
the $A_{1u}$ representation of $O_h$ and that multiplied by
$\alpha_2$ belongs to the $A_{2u}$ representation of $O_h$; both
these vectors map to the representation $A_1$ of $T$. Our result
that the spin-triplet pairing vector $\vd_{\vsk}$ should be
parallel to $\vg_{\vsk}$ for spin-triplet superconductivity to be
stable is unchanged. Consequently, $p$-wave superconductivity is
suppressed for MnSi if $\alpha_1 \lesssim \alpha_2$. The relative
size of $\alpha_1$ and $\alpha_2$ will require band structure
calculations to determine.

We are grateful S. Cunroe,  A. Rosch and  I.A. Sergienko for
useful communications regarding MnSi.


\begin{references}
\bibitem{comment1} This discussion is different from the usual
comparison of superconducting condensation energy and the magnetic
energy (via spin susceptibility), which defines a first order
transition. Our present discussion addresses the situation when the
continuous transition point lies at higher fields than the first
order transition. Note also that large values of $ \alpha \gg k_B T_c
$ yield a drastic enhancement of the low-temperature spin
susceptibility for ``spin-singlet'' pairing states \cite{gor01,yip02}.
\end{references}
\end{document}